\documentclass[preprint]{aastex}

\shortauthors{Weatherall}
\shorttitle{Relativistic-Plasma Compton Maser}

\begin{document}

\title {A Relativistic-Plasma Compton Maser}

\author{James C. Weatherall}
\affil{Department of Physics \\ 
New Mexico Institute of Mining and Technology \\
Socorro  NM 87801}

\begin{abstract}
A relativistic pair-plasma which contains a high excitation of
electrostatic turbulence could produce intense radiation 
at brightness temperature in excess of $10^{20} \; K$ by 
stimulated scattering.  Important relativistic effects 
include the broadband frequency response of the plasma, 
and Compton-boosting of the scattered radiation.  In 
radio-frequency relativistic plasma, the optical depth can 
be as small as tens of meters.  
When the plasma wave excitation is one-dimensional and particle
distributions have $T_{\perp}<< T_{\parallel}$, the frequency-dependent 
angular distribution of the emission exhibits characteristics of pulsar
emission.
\end{abstract}


\section{Introduction}

Radio emission at extremely high brightness temperature
is possible by stimulated scattering in astrophysical plasmas with 
a high degree of plasma turbulence.  Intensities on the order of 
$10^{20}$ to $10^{30}$ K are suspected in some astrophysical objects, 
and an emission process of the form of a plasma maser may explain 
the extraordinary intensity at radio frequencies of pulsars \citep{mel96} 
and active galactic nuclei \citep{wag96}.  
The following calculation exhibits the intensity and spectrum of 
emission in a relativistic plasma of electrons and positrons 
under the assumption of a uniform and
high degree of excitation of plasma electrostatic wavemodes.  It shows 
that the path length through the plasma can be relatively small for 
radiative growth; that Compton-like scattering increases the mean 
emission frequency above the plasma frequency; and the frequency 
bandwidth of the emission is fairly narrow despite the broadband 
excitation in the plasma.

The calculation is done in a relativistic regime because pair plasmas 
derive from extremely energetic processes such as
gamma-ray annihilation.  Simulations of pair plasma creation \citep{are01} 
show that the pair-plasma distribution functions can be described by a 
thermal parameter which is moderately relativistic.  In this paper, the 
kinetic temperature parameter $\rho = mc^2/(k_B T_K)$ is assigned a value 
of 1/10.

Conversion into electromagnetic modes by scattering on plasma waves in 
nonrelativistic space plasma is known to produce radiation at the plasma 
frequency $\omega_p= (4 \pi n e^2/m_e)^{1/2}$, and wave-wave
coalescence at twice the plasma frequency \citep{gur81}.  Radiation at 
these well-defined frequencies is due to the narrow frequency response 
in the plasma.  It seems obvious that emission in a relativistic plasma is
unlikely to follow this plasma emission paradigm.  For one thing, 
electrostatic modes in a relativistic plasma exist over a broad range of 
frequencies, $\omega_p \sqrt{\rho} < \omega <  \omega_p /\sqrt{\rho}$ 
\citep{god75a,god75b,mel00}.
Furthermore, scattering by relativistic particles can modify frequencies
by factors of $4 \gamma^2$.  Still, the stimulated emission appears to
be fairly narrow in frequency, but this is a characteristic of maser 
emission due to frequency-dependent growth acting over many growth lengths.

Although this solution is illustrative of relativistic effects in 
plasma emission, several assumptions are used to simplify calculation.
First, the excitation of the turbulence is characterized by a single 
temperature parameter without regard to specific plasma-streaming or 
shock-excitation mechanisms, or wave cascades.  For example, in thermal 
equilibrium the energy density in plasma waves relative to kinetic energy 
density, $E^2/(8 \pi n k_B T_K)$, is inversely proportional to the
number of particles in a Debye cube,  
$n \lambda^3_D=n[c/(\omega_p \sqrt{\rho})]^3$.  
This parameter can be 
quite large for astrophysical plasmas.  If a radio-frequency plasma 
($\nu_p \sim 3 GHz$) manages an equipartition between turbulent 
electrostatic energy and thermal kinetic energy at temperatures of 
$T_K \sim 5 \times 10^{10} K$, the characteristic excitation as described 
by an enhanced temperature $T_{NL}= (n \lambda^3_D) T_K$ could be more 
than $10^{22} \; K$. This coherent enhancement of scattering on plasma 
waves was recognized by \citet{gal64}, and \citet{col70}.

Another simplification is to do the calculation without an imposed background
magnetic field.  This justifies the assumption of isotropy in the particle 
distributions, wave spectra, and wave dispersion.  Including a magnetic field
would tend to make the turbulence one-dimensional in the 
direction of the magnetic field, and would modify the dispersion 
properties of the electromagnetic modes.  Thus the plasma maser in a 
magnetized plasma would not be istropic in direction, but would have 
preferred directions for emission.
The effect of making the maser plasma one-dimensional is illustrated
by an example.

Finally, the turbulent excitation is assumed to be steady.  This energy 
reservoir in turbulence must be maintained by an injection process against
the radiation loss, but to model this requires an additional
kinetic model for the turbulence.  The assumption of constant wave excitation 
can be expected to fail when the brightness temperature greatly exceeds the 
wave temperature, and radiation becomes a significant energy sink.  
Radiative losses will also determine the lifetime of the system.

The calculation proceeds from classical scattering rates between
plasma and electromagnetic wave modes.  The kinetic equation is
put into the form of a radiative transfer equation in the next
section, assuming a kinetic temperature for the plasma and a nonlinear
effective temperature for the plasma turbulence.  The necessary integrals 
for the scattering coefficients are done numerically as described in
section 3.  The  intense emission which can derive from the
maser process is discussed in section 4.

\section{Transfer Equation}

We start with a transfer equation between the plasma waves and
electromagnetic radiation (see, for example, \citet{mel80}, section 5.4).  
For this equation, the electromagnetic wave spectra is 
described by the photon number density function, $N^T({\bf k})$, and
the plasma wave spectra is described by the number density function
$N^L({\bf k})$.  (The number density spectrum is also called the occupation
number.) 

By definition, the number of photons with wavevector ${\bf k}$ in a 
phase space volume $d^3 k$ around ${\bf k}$ is given by 
$N({\bf k}) \left[d^3k/(2\pi)^3 \right]$.  The integral of $N({\bf k})$ 
over phase space will give the number of photons per $cm^{3}$.

The plasmon occupation number can be formulated from the electric field
energy spectrum of the turbulent waves,
\begin{equation}
W = \frac{1}{VT} \int \frac{E({\bf r},t) E({\bf r},t)}{4 \pi} dV dt 
=\frac{1}{VT} \int \frac{d^{3}k}{(2 \pi)^{3}} \frac{d \omega}{2\pi} 
\frac{E^{*}({\bf k},\omega) E({\bf k},\omega)}{4 \pi} 
\hspace{.2in}
\end{equation}
averaged over time $T$ and volume $V$.
Assuming well-defined wavemodes, 
$E({\bf k},\omega)= E({\bf k}) 2 \pi \delta[\omega- \omega^L({\bf k})]$, 
the integration over $\omega$ gives the following:
\begin{equation}
W= \int \left[ \frac{1}{V} \frac{E^*({\bf k}) E({\bf k})}{4 \pi} \right] \; 
\left[ \frac{d^3k}{(2 \pi)^3} \right]
\hspace{.2in}.
\end{equation}
Because the term in second  brackets is the number density of wavemodes
between ${\bf k}$ and ${\bf k} + d{\bf k}$, it is easy to identify the 
first term in brackets as the electrostatic energy per wavemode.  We 
apply a thermal-like excitation of normal modes in the turbulence to derive:
\begin{equation}
\frac{1}{V} \frac{E^{*}({\bf k}) E({\bf k})}{4 \pi} = \frac{1}{2} k_{B} T_{NL}
\hspace{.2in}.
\end{equation}
For electrostatic waves, another degree of freedom is invested in
the particle motion.
Finally, the occupation number is acquired from a semi-classical formula
\begin{equation}
N^L({\bf k}) = 
\cases{ 
\frac{k_B T_{NL}}{\hbar \omega^L(k)} ,   
& $\sqrt{\rho} \; {\omega}_p < kc <   {\omega}_p /\sqrt{\rho} \hspace{.2in}$; 
\cr
0 ,  &  otherwise. } 
\end{equation}
The inequality describes the frequency range of plasma normal modes
in a relativistic plasma.

With this notation, the transfer equation, in the classical limit,
for photons in plasma turbulence is given by (\citet{mel80}, eq. 5.89):
\begin{eqnarray}
\frac{dN^T({\bf k})}{dt} &=& \int \frac{d^3k'}{(2 \pi)^3} d^3p 
 n \, \varpi^{TL}({\bf p},{\bf k},{\bf k}') 
\nonumber \\ &\mbox{ }&  \times
\left[ f({\bf p}) \left\{N^L({\bf k}') - N^T({\bf k}) \right\} 
+ N^L({\bf k}') N^T({\bf k}) \; \hbar({\bf k}-{\bf k}') 
\cdot \frac{\partial f}{\partial {\bf p}} \right] .
\end{eqnarray}
The three-dimensional thermal distribution of lepton momentum in a 
relativistic plasma is given by the normalized function 
$f({\bf p}) d^3p = exp(-\rho \gamma(p))/Z \; d^3p$; 
here, $Z= (4 \pi K_2(\rho))/\rho$, 
$K_2$ is the modified Bessel function, $\gamma = [1+p^2/(m^2c^2)]^{1/2}$,
and $\rho$ is the temperature parameter.  The spatially-averaged number 
density of particles in the zero-momentum reference frame is $n$.
To write this equation in terms of the specific intensity $I(\omega)$ for
a single polarization, use
\begin{equation}
I(\omega) d \omega d \Omega = \hbar \omega N^T(\omega/c) c 
\frac{ k^2 dk d \Omega}{(2 \pi)^3}
\hspace{.2in}.
\end{equation}
The resulting equation has three terms.  The first term represents 
spontaneous scattering:
\begin{equation}
\frac{dI}{ds} = \Lambda_1(\omega) \frac{\omega^2 }{(2 \pi)^3 c^3} k_BT_{NL}
\hspace{.2in},
\end{equation}
where the emission coefficient has been put into the form
\begin{equation}
\Lambda_1(\omega) =  
\int \frac{d^3k' }{(2 \pi)^3} n \,f({\bf p})  d^3p 
\varpi^{TL}({\bf p},{\bf k},{\bf k}') 
\frac{\omega}{\omega'}
\hspace{.2in}.
\end{equation}
The second term represents absorption
\begin{equation}
c\frac{dI}{ds} =  -  \Lambda_2 (\omega) I(\omega)
\hspace{.2in},
\end{equation}
where the absorption coefficient is
\begin{equation}
\Lambda_2(\omega) =  
 \int \frac{d^3k' }{(2 \pi)^3}  n f({\bf p})  d^3p 
\varpi^{TL}({\bf p},{\bf k},{\bf k}') 
\hspace{.2in}.
\end{equation}
The final term contributes to stimulated scattering.  For a thermal 
distribution, the momentum derivative gives
\begin{equation}
\frac{\partial f}{\partial {\bf p}} = -\frac{c f({\bf p})}{k_BT} \; 
 {\bf \beta}
\hspace{.2in}.
\end{equation}
The vector arithmetic can be worked out using the kinematic relationship
between frequencies,
\begin{equation}
\omega'= \omega \frac{1 - \beta \cos \theta}{1- \beta \cos\theta'}
\hspace{.2in},
\end{equation}
where the angles are between the wavevectors and the electron velocity 
vector.  Thus,
\begin{equation}
\hbar ({\bf k}-{\bf k}') \cdot \frac{\partial f}{\partial {\bf p}} = 
-\frac{\hbar \omega}{k_BT} 
\left[ 1 - \frac{\omega'}{\omega} \right] \, f({\bf p})
\hspace{.2in}.
\end{equation}
The stimulated scattering term is
\begin{equation}
c \frac{dI}{ds} =  \frac{T_{NL}}{T}
\left[ \Lambda_2(\omega) - \Lambda_1(\omega) \right] I(\omega)
\hspace{.2in}.
\end{equation}
The complete transfer equation is
\begin{equation}
\frac{dI(\omega)}{ds} =  \Lambda_1 \frac{\omega^2 k_BT_{NL} }{(2 \pi)^3 c^3}  
+ \left[ -\Lambda_2 + \frac{T_{NL}}{T} \left( \Lambda_2 - \Lambda_1 \right) 
\right] \frac{I(\omega)}{c}
\hspace{.2in}.
\end{equation}

\section{Formulation of the Scattering Coefficients}

The probability for scattering of longitudinal
waves into transverse waves by relativistic electrons is given by
(\citet{mel80}, eq 4.150; also \citet{mel71})
\begin{eqnarray}
\label{probability}
& &n \varpi^{TL}({\bf p},{\bf k},{\bf k}')  \frac{d^3k' }{(2 \pi)^3}  =  
\frac{(2 \pi)^3 n e^4}{m^2 \omega' \omega} 
\frac{(1-\beta^2)}{(1-\hat{k} \cdot {\beta})^2 
(1 - \hat{k}' \cdot {\beta})^2} \;
\delta \left( \omega (1 - \hat{k} \cdot {\beta}) 
- \omega' (1- \hat{k}' \cdot {\beta}) \right) 
\nonumber \\ 
 & &  \mbox{ } \times     
\left[ (1  -  \hat{k} \cdot {\beta})^2 \left( 1 
- (\hat{k}' \cdot {\beta})^2 \right) -
(1-\beta^2) (\hat{k} \cdot \hat{k}' - \hat{k}' \cdot {\beta})^2 \right] 
\frac{d^3k' }{(2 \pi)^3} 
\hspace{.2in}, 
\end{eqnarray}
where use is made of $\omega' = k' c$.  
$n \; \varpi^{TL}({\bf p},{\bf k},{\bf k}') \; d^3k'$ 
has units $s^{-1}$.  
In Equation (\ref{probability}), a sum is made over polarization in the 
scattered waves.

The integration of the scattering rates over $d^3k'$ and $d^3p$ are done 
in spherical coordinates in which angle $\theta'$ measures ${\bf k}'$ relative
to ${\beta}$ and $\theta$ measures ${\beta}$ relative to ${\bf k}$.
In terms of angle cosines $\mu= \cos{\theta}$:
\begin{eqnarray}
d^3k' &=& \frac{\omega'^2}{c^3} d\omega' \, d\phi' \, d\mu' 
\hspace{.2in}; \nonumber \\
d^3p &=& p^2 dp d \phi d\mu 
\hspace{.2in}.
\end{eqnarray}
The integral over $d\omega'$ can be done easily with the delta-function.
After integration over $d\phi'$ and $d \phi$,
the remaining integrals over
angle in the emission rate are
\begin{equation}
\label{lambda1}
\Lambda_1(\omega) = \frac{3}{16} n \sigma_T c \int 4 \pi \; p^2 f(p) dp 
\int_{\mu_{min}}^{1} \frac{1-\beta^2}{1 - \beta \mu} d \mu
\int_{-1}^{\mu'_{max}} \frac{A + B \mu'^2}{(1-\beta \mu')^4} d \mu'
\hspace{.2in},
\end{equation}
where $A$ and $B$ are algebraic functions of $\mu$:
\begin{eqnarray}
A &=& 1 + \beta^2 -4 \beta \mu + \mu^2 + \beta^2 \mu^2 
\hspace{.2in}, \nonumber \\
B &=& 1 - 5 \beta^2  + 2 \beta^4  + 4 \beta \mu  -3 \mu^2  + 
3 \beta^2 \mu^2  -2 \beta^4 \mu^2  
\hspace{.2in}.
\end{eqnarray}
The upper cutoff to the $d\mu'$ integral is a byproduct of the delta-function,
and the high-frequency bound on $\omega' < (1+\beta) \gamma \omega_p$.  Thus,
\begin{equation}
\mu'_{max} = min  \cases{ 
\frac{1}{\beta} -\frac{\omega(1-\beta \mu)}{2 \beta \gamma \omega_p} 
, \cr
1 .
}
\end{equation}
In order for $\mu'_{max}$ to be larger than $-1$, the angles $\mu$ 
must be larger than
\begin{equation}
\mu_{min} = max  \cases{  
\frac{1}{\beta} - \frac{1+\beta}{\beta}  
\frac{2 \gamma \omega_p}{\omega} , \cr 
-1 .
}
\end{equation}
Similarly, the absorption rate
\begin{equation}
\label{lambda2}
\Lambda_2(\omega) = \frac{3}{16} n \sigma_T c \int 4 \pi \; p^2 f(p) dp 
\int_{\mu_{min}}^{1} \frac{1-\beta^2}{(1 - \beta \mu)^2} d \mu
\int_{-1}^{\mu'_{max}} \frac{A + B \mu'^2}{(1-\beta \mu')^3} d \mu'
\hspace{.2in}.
\end{equation}

The integrals in the above equation are completed numerically. Solutions  
for $\Lambda_1(\omega)$ and $\Lambda_2(\omega)$ are shown
in Figure 1.  The two scattering rates are equal 
($\Lambda_1 \sim \Lambda_2$) near the characteristic
frequency $\omega_p (1/\rho^{3/2})$.

\section{Amplification by Stimulated Emission}

With fixed values
for the wave temperature, $T_{NL}$, and the kinetic temperature, $T_K$, 
the transfer equation is a first order differential
equation with constant coefficients which can be solved directly as a 
function of path length:
\begin{equation}
I = I_K e^{s \Lambda_{stim}/c} + I_{NL} \frac{\Lambda_1}{\Lambda_{stim}} 
\left( e^{s \Lambda_{stim}/c} -1 \right)
\hspace{.2in},
\end{equation}
where the effective stimulated scattering rate is
\begin{equation}
\Lambda_{stim}(\omega)= (\Lambda_2 - \Lambda_1) T_{NL}/T_K - \Lambda_2
\hspace{.2in}.
\end{equation}

At low frequency ($\omega \ll  \omega_p \rho^{3/2}$), $\Lambda_1 
\gg \Lambda_2$ and the intensity is limited to the kinetic temperature.  
Near the characteristic frequency, the intensity saturates at the nonlinear 
wave temperature.  At higher frequencies, the intensity increases
exponentially.  However, the exponentiation rate diminishes at frequencies
much higher that the characteristic frequency because $\Lambda_1$ and 
$\Lambda_2$ become small.   

The high brightness temperatures derived from this theory must be qualified 
by the fact that cooling of the plasma and the turbulence are not 
taken into account.  A full picture of the energy balance requires 
additional kinetic equations for the plasma waves and particle energy 
distribution.  However, it is easy to estimate how long the plasma can 
maintain its relativistic energy against the Compton loss.  
At $10^{22} K$, a $10^{3} cm$ scale system 
(such as a pulsar magnetospheric source region) will last $10^{-6} s$, 
and a $10^{14} cm$ scale system (encompassing an AGN accretion region)
will last for $10^{5} s$ against radiative loss.  
These simple numbers suggest how maser lifetime might relate to 
microstructure in pulsar radio emission and intraday 
variability in QSO's.

To show the magnitude of stimulated emission, the scattering coefficients 
can be scaled as follows:
\begin{equation}
\frac{s \Lambda}{c} \; \frac{T_{NL}}{T_K} =
1.3 \times 10^{-10} \; \frac{T_{NL}}{T_K} \; 
\left( \frac{\Lambda}{n \sigma_T c} \right) 
\left( \frac{n}{10^{11} \; cm^{-3}} \right)  \; 
\left(\frac{s}{1000 \; cm}\right)  
\hspace{.2in}.
\end{equation}  
Substantial growth in intensity occurs in path lengths smaller than a
kilometer assuming turbulence temperatures $T_{NL}/T_K \ge 10^{12}$, 
as demonstrated in Figure 2.     
Such a large temperature for the turbulence is not implausible from an energy
standpoint.  The turbulent temperature estimated from equipartition
 between electrostatic and plasma kinetic energy is
\begin{equation}
\frac{T_{NL}}{T_K} \sim \left( \frac{c \sqrt[3]{n}}{\omega_p \sqrt{\rho}} \right)^3 =
6 \times 10^{13} \; \left( \frac{10^{11} \; cm^{-3}}{n} \right)^{1/2}
\; \left( \frac{0.1}{\rho} \right)^{3/2}
\hspace{.2in}.
\end{equation}
For a plasma kinetic temperature on the order of $10^{10} \; K$, 
radiative brightness temperatures greater than $10^{23} K$ are consistent 
with stimulated emission from this mechanism.  
Higher brightness temperatures are possible with greater turbulent 
temperature or larger path lengths, depending on what limits are
imposed by Compton cooling. 

In summary, the relativistic-plasma Compton maser uses free energy
in the form of wave turbulence to produce electromagnetic radiation 
via induced scattering in the relativistic thermal plasma.  The high 
brightness temperature derives from the large value of the plasma parameter,
$n[c/(\omega_p \sqrt{\rho})]^3$ -- generally true in astrophysical
plasmas, although less so in laboratory plasmas.
The maser turbulence conversion 
mechanism is an alternative to other plasma turbulence conversion processes 
invoking coherent spatial effects or nonlinear waves dynamics (for example, 
\citet{wea97,wea98,ass90}), which may not develop due to the turbulence 
being strongly driven or highly inhomogeneous. 

The turbulent conversion process described here might also apply to
pulsars.  Maser models are not new to pulsar radiation physics (for example,
\citet{lyu99,luo95}), but these masers resemble free-electron masers 
in which particle beams generate the emission.
A model invoking the Compton plasma-maser requires further inclusion of
anisotropies due to the magnetic field, wave dispersion properties in
magnetized plasma, and specific mechanisms for wave excitation by
collimated particle beams.
However, we can simulate these effects by limiting the plasma momentum
distribution and turbulent wavevectors to a single coordinate axis. 

In one-dimension, the differential scattering cross-sections in terms
of the angle cosine $\mu$ between the emission wavevector ${\bf k}$ and the
$z$-coordinate axis are given by
\begin{eqnarray}
\frac{d \Lambda_2(k,\mu)}{d \Omega}  &=& 
\int \frac{dk'}{(2 \pi)^3} \frac{\omega'^2}{c^2} \; \int n dp f(p) 
\; \omega^{TL}(p,k',{\bf k}) \nonumber \\
\frac{d \Lambda_1(k,\mu)}{d \Omega}  &=& 
\int \frac{dk'}{(2 \pi)^3} \frac{\omega'^2}{c^2} \; \int n dp f(p) \; 
\left[\frac{1 \pm \beta}{1-\beta \mu} \right] \; \omega^{TL}(p,k',{\bf k})
\end{eqnarray} 
 The choice of sign is $(+)$ for electrostatic waves with $k'$ in the
{\it negative} $z$-direction, and $(-)$ for waves in the {\it positive}
$z$-direction.  The calculation assumes waves have relativistic 
dispersion relation $\omega'=k'c$, and the uniform excitation applies to
frequencies $\omega_p \, \sqrt{\rho}  < \omega' < \omega_p /\sqrt{\rho}$.
The one-dimensional electron distribution function is given by
$
f(p) \; dp =  \exp\left[{-\rho \gamma(p)}\right] \; 
dp/\left({2 K_1(\rho)}\right)
$,
where $K_1$ is the modified Bessel function.

The transfer equation is the same as before, substituting for
 the scattering
rates   $\Lambda_1 \rightarrow \Delta \Omega (d \Lambda_1/d \Omega)$
and $\Lambda_2\rightarrow \Delta \Omega (d \Lambda_2/d \Omega)$.
Here, $\Delta \Omega$ is the range of solid angles about the turbulence
axis which have the nonlinear excitation.  
The important difference in one-dimension is the directivity of the maser.  
Figure \ref{polar} is a polar plot in angle and frequency.  The
emissivity in the plasma rest frame is largest in directions 
{\it near} perpendicular to the magnetic
axis, with some angular structure due to relativistic thermal velocities.
However, the plasma is moving relative to the lab/star 
frame because of the polar cap current flow, and  
the emission can be expected to be beamed relativistically: this is 
illustrated in the figure with a Lorentz transformation.

One intriguing consequence of the moving maser is the angular dependence
of the peak emission for different frequencies.  Radio pulsars show a variation
in pulse profile (intensity vs. phase) for different observing
frequency.  This is generally interpreted as radius-to-frequency mapping based
on the presumption that the emission comes from different heights in the polar
cap, and that the emission frequency is tied to the local plasma frequency. 
As clearly shown in Figure \ref{phase}, the maser emission from a single
location can produce similar frequency-dependent profiles.  Note that the
spectrum of the emission is found here to have a fairly steep index in the 
case of uniform one-dimensional turbulence, $I_{\nu} \sim \nu^{-3}$.   

Finally, we remark that relativistic temperatures have the 
effect of supporting wave frequencies which are in a broad range about
the plasma frequency: still, the highest growth occurs at frequencies above
the plasma frequency.  Thus, relativistic effects do not appear to mitigate 
the puzzle that emission at the local plasma frequency in the pulsar 
polar cap plasma produces frequencies which seem high for radio emission 
\citep{mel2K,kun98}.

\acknowledgments 

This work is supported by NSF grants AST-9618408 and AST-9720263.  
Justin Jayne contributed to evaluating the numerical integrations. 
Discussions with Paul Arendt, Jean Eilek, and Tim Hankins are gratefully 
acknowledged.

\clearpage
%figure 1
\begin{figure}[4 in]
\plotone{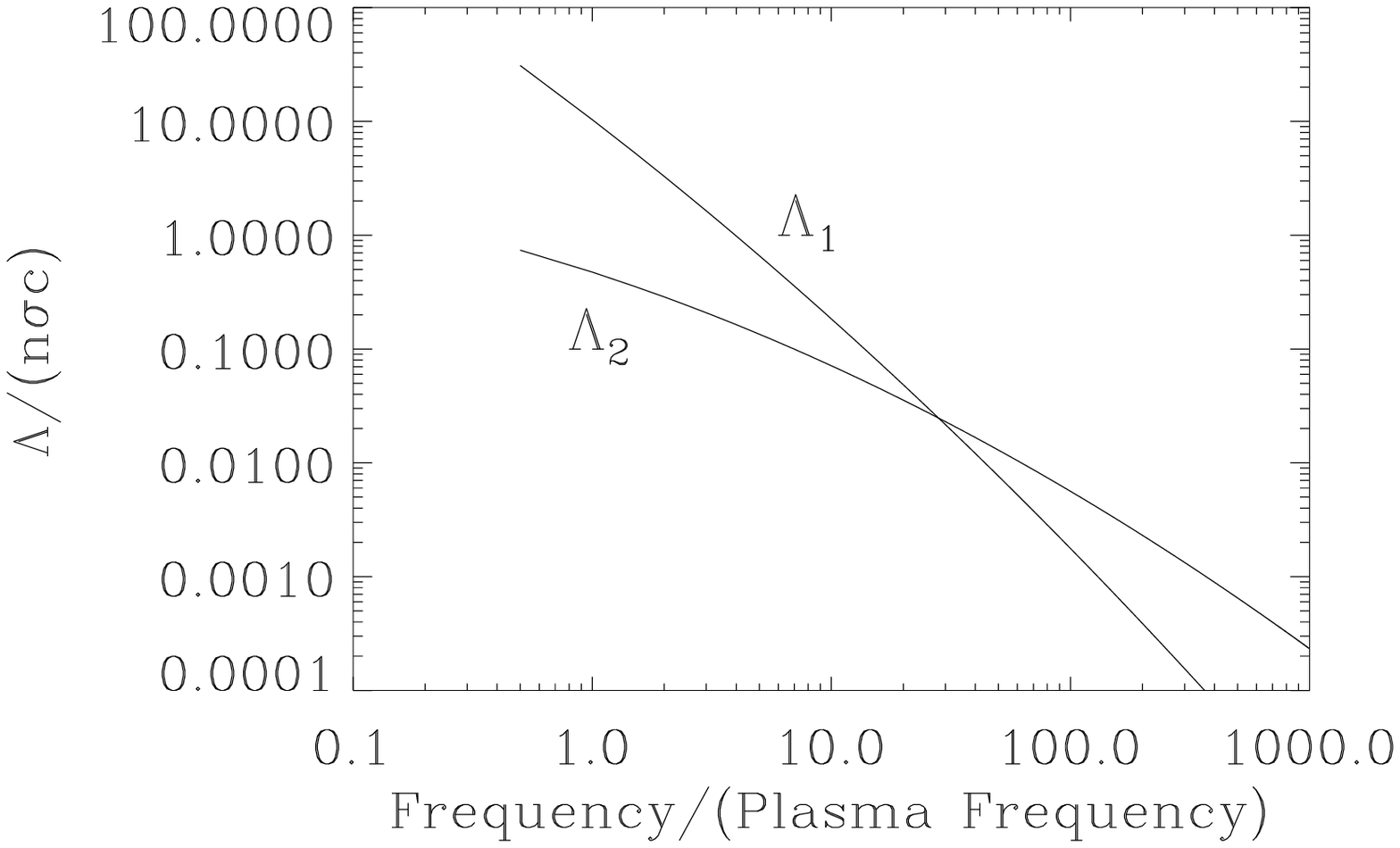}
\caption[]{\label{lambdas}
Scattering coefficients determined numerically from 
Eqs. \ref{lambda1} and \ref{lambda2} for isotropic thermal 
particles of temperature $k_BT_K=10mc^2$.}
\end{figure}

\clearpage

%figure 2
\begin{figure}
\plotone{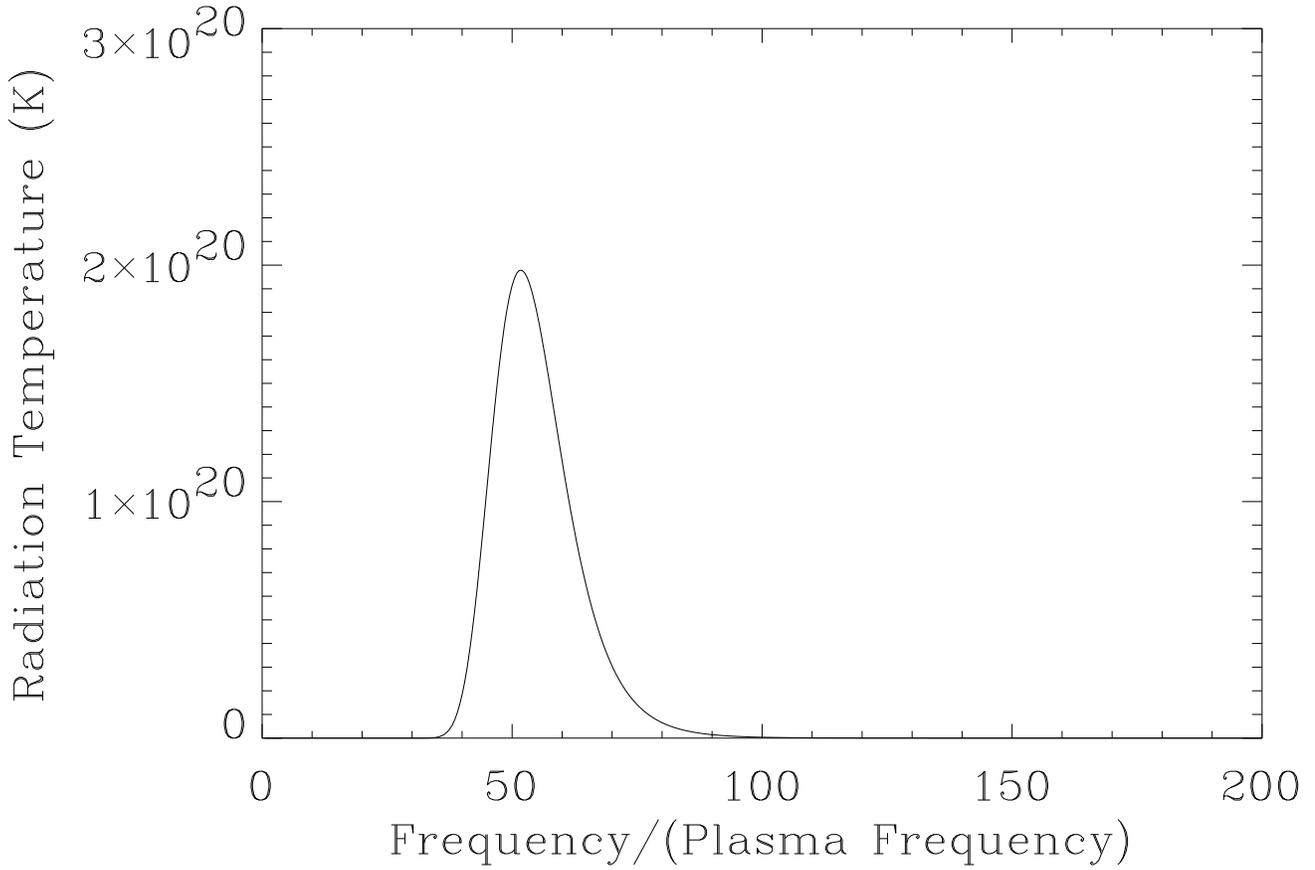}
\caption[]{\label{spectrum}
Intensity of plasma maser emission in a electron-positron plasma
with kinetic temperature of $10 mc^2$.  The turbulent excitation is
taken to be enhanced by $T_{NL}/T_K =  10^{12}$.  Scaling to a
plasma frequency of $f_0 = 3 GHz$, the path length
corresponds to $s= 2 \times 10^5 \; cm$.   
  } 
\end{figure}
\clearpage

%figure 3
\begin{figure}
\plotone{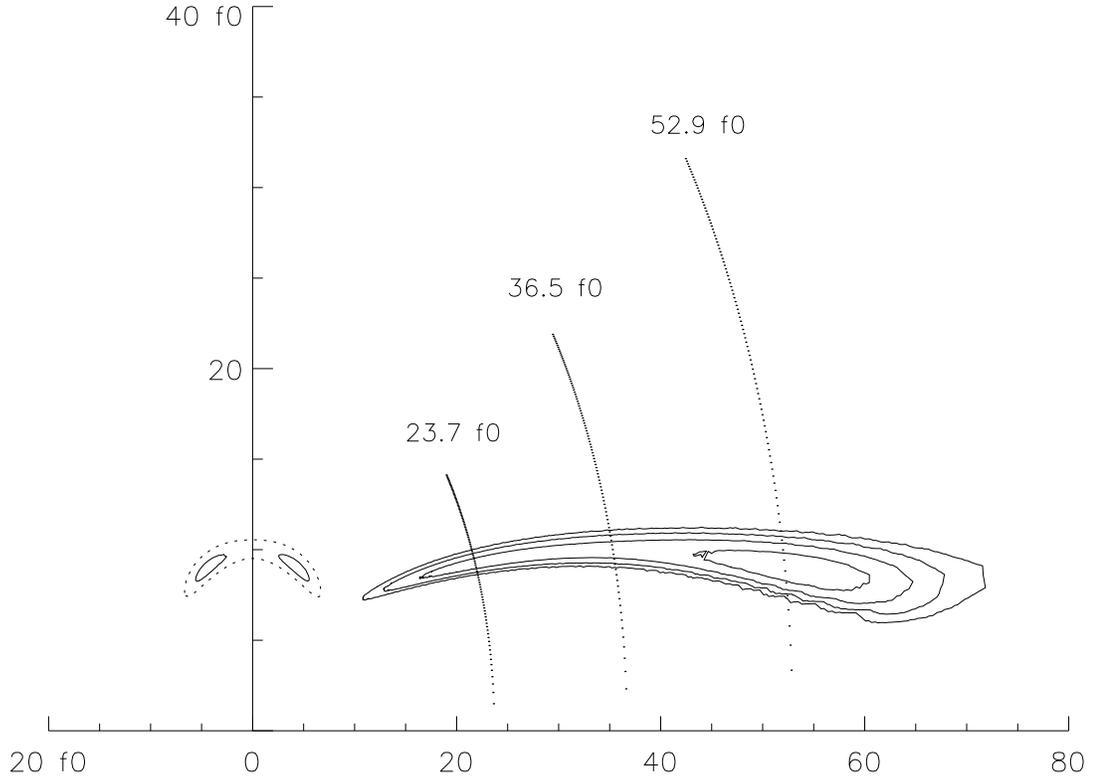}
\caption[]{\label{polar}
Contours of constant brightness temperature in frequency space for
the Compton-maser for one-dimensional turbulence with $T_{NL}/T_{K}=10^{12}$,
and a path length of $s= 1.5 \times 10^6 \; cm$.  The frequency is in 
units of the plasma frequency, $f_0$.  The contour levels are 
$1, \; 2, \; 4,$ and $8 \times 10^{16} \; K$.  The contours for both the 
plasma rest frame, and the lab frame in which the plasma moves with 
$\gamma=3.8$ are shown: the lowest contour is dotted in the former case.  
Intensity along the three cuts of constant frequency are presented in
the next figure.
}
\end{figure}
\clearpage
%figure 4
\begin{figure}
\plotone{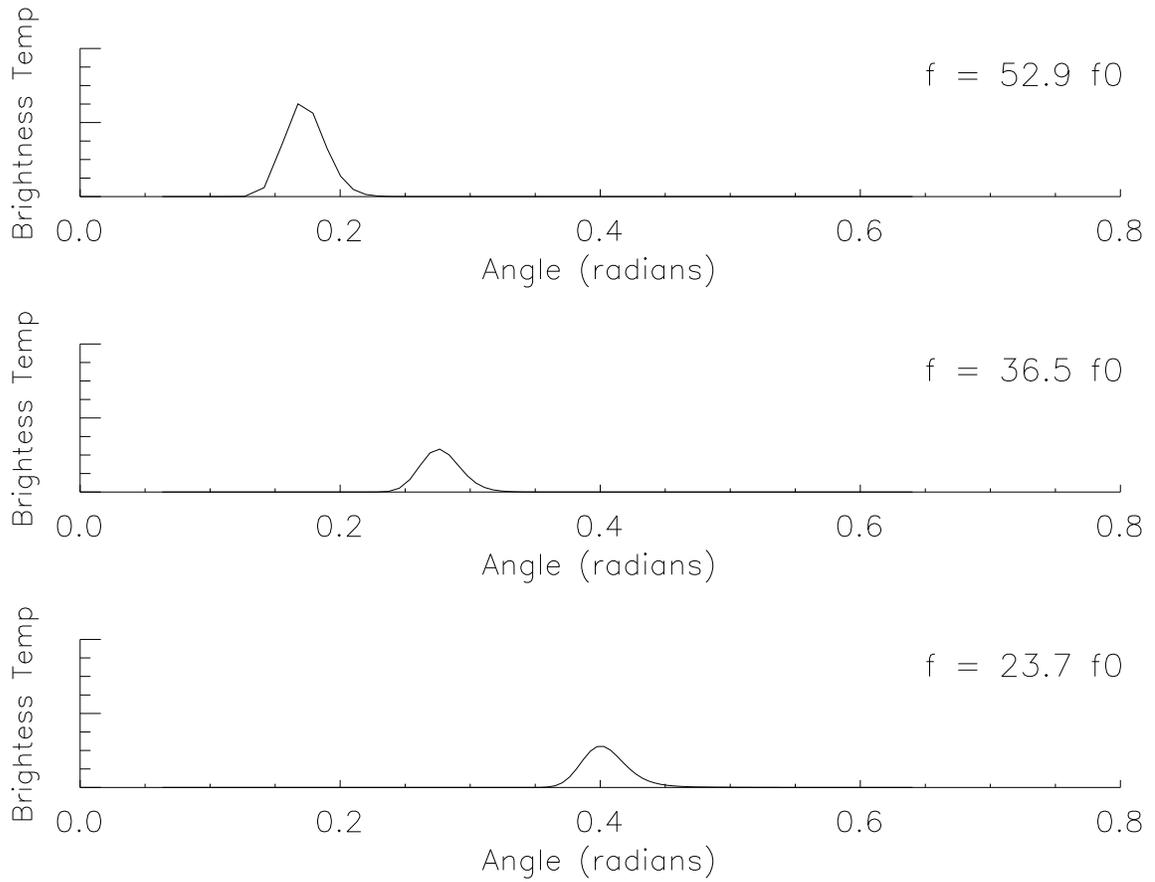}
\caption[]{\label{phase}
Brightness temperature vs. angle for three different frequencies for the
case of a moving plasma.  The temperature scale is the same for each plot,
with a scale maximum at $1.5 \times 10^{17} \; K$.
}
\end{figure}

\end{document}